\documentclass[11pt,twoside]{article}

\usepackage{asp2006}
\usepackage{epsf}
\usepackage{psfig}
\usepackage{lscape}

\begin{document}
\title{Were All Massive Stars Born in OB Associations and Clusters?}  
\author{You-Hua Chu and Robert A. Gruendl}
\affil{Astronomy Department, University of Illinois,
1002 West Green Street, Urbana, Illinois 61801, USA}

\begin{abstract} 
It has been commonly conjectured that all massive ($>$ 10 $M_\odot$)
stars are born in OB associations or clusters.  Many O and B stars in 
the Galaxy or the Magellanic Clouds appear to exist in isolation, 
however.  While some of these field OB stars have been ejected from 
their birthplaces, some are too far away from massive star forming
regions to be runaways.  Can massive stars form in isolation?  The 
{\it Spitzer} survey of the Large Magellanic Cloud (aka SAGE) provides
a unique opportunity for us to investigate and characterize the formation
sites of massive stars for an entire galaxy.  We have identified all
massive young stellar objects (YSOs) in the Large Magellanic Cloud.  
We find that $\sim$85\% of the massive YSOs are in giant molecular 
clouds and $\sim$65\% are in OB associations.  Only $\sim$7\% of the
massive YSOs are neither in OB associations nor in giant molecular clouds.
This fraction of isolated massive stars in the Large Magellanic Cloud
is comparable to the 5--10\% found in the Galaxy.
\end{abstract}


\section{Introduction} 
Inventories of massive O-type stars in the Galaxy show that
$\sim$70\% of them are associated with OB associations or clusters,
and that more than 1/3 of the remaining 30\% are runaway OB stars 
\citep{Gies87,Metal98,Metal04}.  These results imply that less 
than 20\% of Galactic O-type stars were formed in isolation.  

Recently, the origin of 43 field stars from the \citet{Metal98}
sample of 227 O stars with $V <$ 8 has been investigated: deep 
images show that 5 of them are in small clusters \citep{witetal04},
and analysis of their location and space velocities suggests
that 22 are likely runaways \citep{witetal05}.
Only 10--20 of the sample of 227 bright Galactic O stars cannot 
be assigned to any OB associations or clusters, indicating that
5--10\% of the Galactic O stars are truly isolated (Zinnecker 
\& Yorke 2007).

The formation of isolated massive stars is of great interest.
While studies of Galactic O stars have yielded important results
on the fraction of O stars formed in isolation, it is not known 
under what interstellar conditions isolated massive stars are formed
or why field stars have steeper initial mass functions (IMFs) than
OB associations \citep{Metal95}. 
To answer these questions, we need to examine young massive stars 
before their energy feedback has significantly altered the ambient
interstellar conditions and dispersed the natal clouds.
It is also necessary to inventory young massive stars in an entire
galaxy to gain statistical insights.
Recent {\it Spitzer Space Telescope} observations of the Large 
Magellanic Cloud (LMC) provide an ideal dataset to study young
massive stars.

\section{Spitzer Sample of Massive YSOs in the LMC}

The LMC is at a distance of 50 kpc (distance modulus = 18.5),
where 1$''$ corresponds to 0.25 pc.  Its nearly face-on
orientation allows a clear view of the entire galaxy with
little confusion and extinction along the line-of-sight.
{\it Spitzer} observations are able to resolve individual stars
in the mid-IR for the first time \citep{Jetal05,Cetal05}.
We have used our deep IRAC and MIPS observations of seven
LMC H\,II complexes from {\it Spitzer} Cycle 1 and the SAGE
survey of the entire LMC from Cycle 2 \citep{Metal06}
to identify massive young stellar objects (YSOs) in the LMC.

We have retrieved the above mentioned {\it Spitzer} observations 
of the LMC, and carried out photometry for all point sources in 
the four IRAC bands and three MIPS bands.  The MIPS 70 and 160 $\mu$m
data are less useful because of their poor angular resolution.
The SAGE survey of the LMC contains observations made in two epochs.
Photometric measurements were made for each epoch separately, 
compared to reject transients and spurious sources, and combined
to improve signal-to-noise ratios.
To check our method of photometric measurements, we have retrieved 
the {\it Spitzer} Wide-area Infrared Extragalactic (SWIRE) Legacy 
Survey \citep{Letal04} data and made our measurements; 
we find general consistency between our and their results.

\begin{figure}
\centerline{
\psfig{file=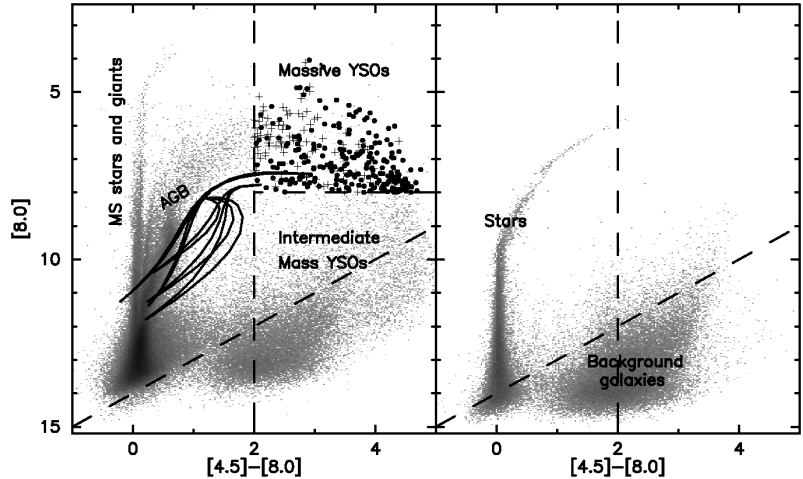,height=2.5in,angle=0}
}
\caption{Color-magnitude diagrams of all point sources in the
LMC (left panel) and the SWIRE survey (right panel).  The massive
YSO candidates in the LMC are located in the upper right corner
of the CMD; the AGB/post-AGB stars are marked by ``+'' and the 
YSOs by filled circles.
}
\end{figure}

The photometric measurements in IRAC bands and MIPS 24 $\mu$m band
were used to assemble various diagnostic color-color and 
color-magnitude diagrams (CCDs and CMDs).  Among these, the most 
informative is the [8.0] vs [4.5]$-$[8.0] CMD.
In Figure 1 we present such CMDs for all point sources in the
LMC (left panel) and in the SWIRE survey (right panel) 
for comparison.
The SWIRE CMD shows a vertical branch of stars of which the
upper part bends to the right owing to the artifact that the 
brightest stars are saturated in the 4.5 $\mu$m band, as well
as a concentration of galaxies.
The CMD of the LMC is much more complex, as it contains normal
stars with zero color, evolved stars with IR excess, YSOs with
IR excess, and background galaxies.

To select YSO candidates in the LMC, we first adopt the galaxy
discriminator suggested by \citet{Hetal06} --  the tilted dashed
line in Figure 1.  As shown in the SWIRE CMD, this discriminator
is indeed an effective upper boundary for most background galaxies.
To exclude normal stars and evolved stars, we adopt a color cutoff
of [4.5]$-$[8.0] $\ge$ 2.0. 
In the LMC CMD, we have overplotted the expected locations of
3000 $L_\odot$ asymptotic giant branch (AGB) and post-AGB stars 
from \citet{Gr06}; it is evident that our discriminator does
exclude most of the normal and AGB/post-AGB stars.
Finally, we adopt the [8.0] $\le$ 8 criterion to select
massive YSOs with masses greater than $\sim$10 $M_\odot$
\citep{Cetal05}.
As show in Figure 1, the massive YSO candidates are in the upper 
right corner of the CMD; however, some AGB and post-AGB stars
are still expected to occupy this region.  

\begin{figure}
\centerline{
\psfig{file=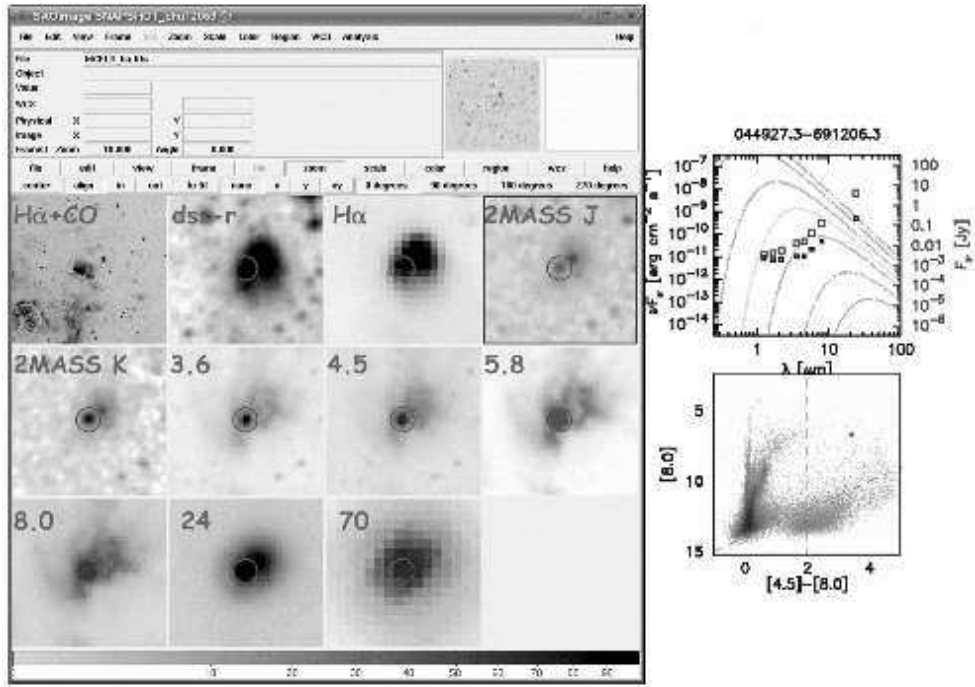,width=5.15in,angle=0}
}
\caption{Determining the nature of the YSO candidate 044927.3$-$691206.3.
Images at multiple wavelengths are displayed simultaneously using ds9.  
The left-most panel in the top row displays an H$\alpha$ image with 
NANTEN CO contours to show the large-scale environment; the other 
panels are matched close-up images from the Digitized Sky Survey red
(dss-r), MCELS H$\alpha$, 2MASS $J$ and $K$, IRAC 3.6, 4.5, 5.8, and 8.0
$\mu$m, and MIPS 24 and 70 $\mu$m bands.  Plots of the YSO candidate's
SED and location in the [8.0] vs [4.5]$-$[8.0] CMD are displayed to the 
right.  These images and plots are examined simultaneously to assess 
the nature of the object.
}
\end{figure}

To determine the true nature of the massive YSO candidates, it 
is necessary to examine their images and spectral energy 
distributions (SEDs) carefully.  
Our approach is illustrated in Figure 2, where ds9 is used to
display a large-scale H$\alpha$ image overplotted with CO
contours, and close-up images in Digitized Sky Survey red (dss2r),
H$\alpha$, 2MASS $J$ and $K$, IRAC 3.6, 4.5, 5.8, and 8.0 $\mu$m,
and MIPS 24 and 70 $\mu$m, alongside plots showing the YSO
candidate's SED and location in the [8.0] vs [4.5]$-$[8.0] CMD.
The H$\alpha$ images are from the Magellanic Cloud Emission Line
Survey (MCELS, Smith et al.\ 1999), and the CO contours are from
the NANTEN survey \citep{Fetal99,Fetal01}.
By examining these images and plots, we can easily identify
foreground stars, AGB/post-AGB stars, and bright galaxies.
After eliminating these contaminating sources, we are left with
234 probable (high-confidence) and 14 possible ($\sim$50\% probability)
massive YSOs.  Details of the {\it Spitzer} photometric measurements
and selection of massive YSOs will be reported by Gruendl \& Chu
(2008, in preparation).

Massive stars are often born in complex environments and the
{\it Spitzer} resolution is inadequate to show the detail.
The 2MASS images are also of limited usefulness.  We have
thus used the Infrared SidePort Imager (ISPI) on the 4m
Blanco telescope at Cerro Tololo Inter-American Observatory
to obtain deeper $J$ and $K$ images with higher angular resolution.
The necessity of such high-resolution images is illustrated in
Figure 3, where close-up images of the object shown in Figure 2
are displayed.  The near-IR counterpart of this source appear 
unresolved in the 2MASS $J$ and $K$ images, but is resolved into
a diffuse source and two point sources in the ISPI $J$ and $K$
images, with one source being significantly brighter in $K$ than 
in $J$. This example fully demonstrates the danger in blindly 
modeling the SED of a YSO in the LMC for its physical properties
without using high-resolution images to examine the multiplicity
of the YSO.

\begin{figure}
\centerline{
\psfig{file=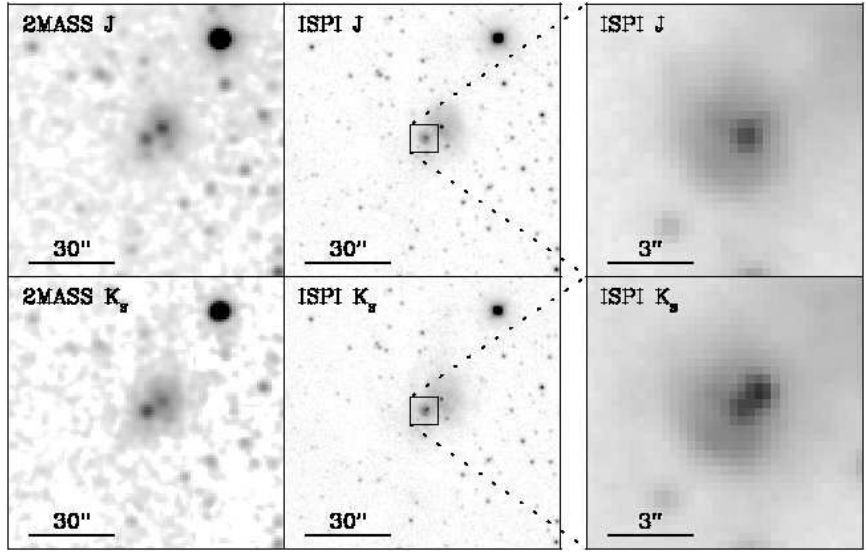,width=4.5in,angle=0}
}
\caption{Near-IR images of the massive YSO candidate
044927.3$-$691206.3 in the LMC.
The YSO is located at the center of the field-of-view.
The two left panels display the 2MASS $J$ and $K$ images,
and the middle two panels ISPI $J$ and $K$ images.
The central region is enlarged and shown in the two
right panels.  The diffuse emission and the two point sources
resolved by the ISPI images appear as a point source in 
2MASS and {\it Spitzer} IRAC images in Figure 2.
}
\end{figure}

\section{Environments of Massive YSOs}

\begin{figure}
\centerline{
\psfig{file=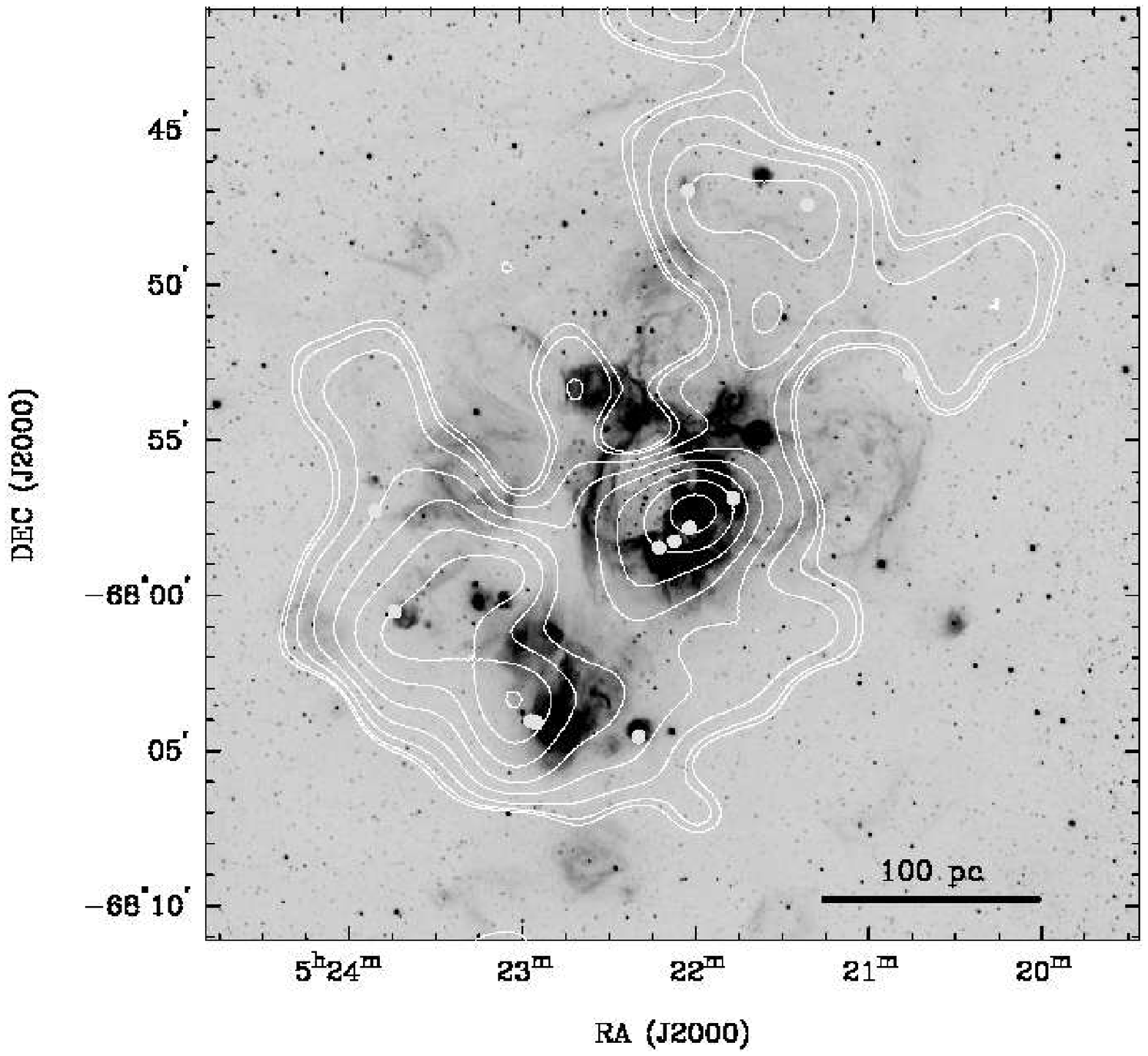,width=2.4in,angle=0}
\psfig{file=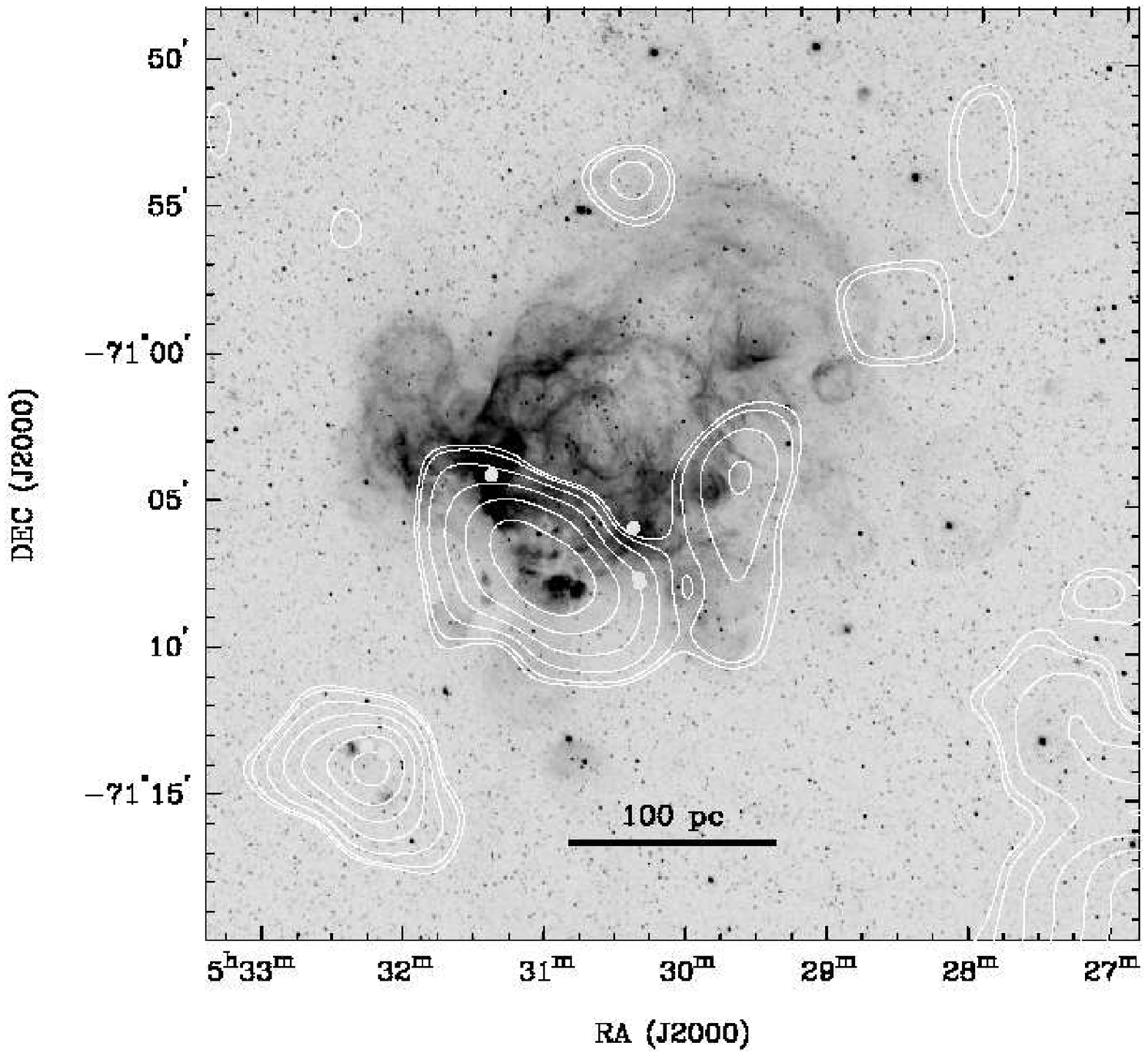,width=2.4in,angle=0} }
\centerline{
\psfig{file=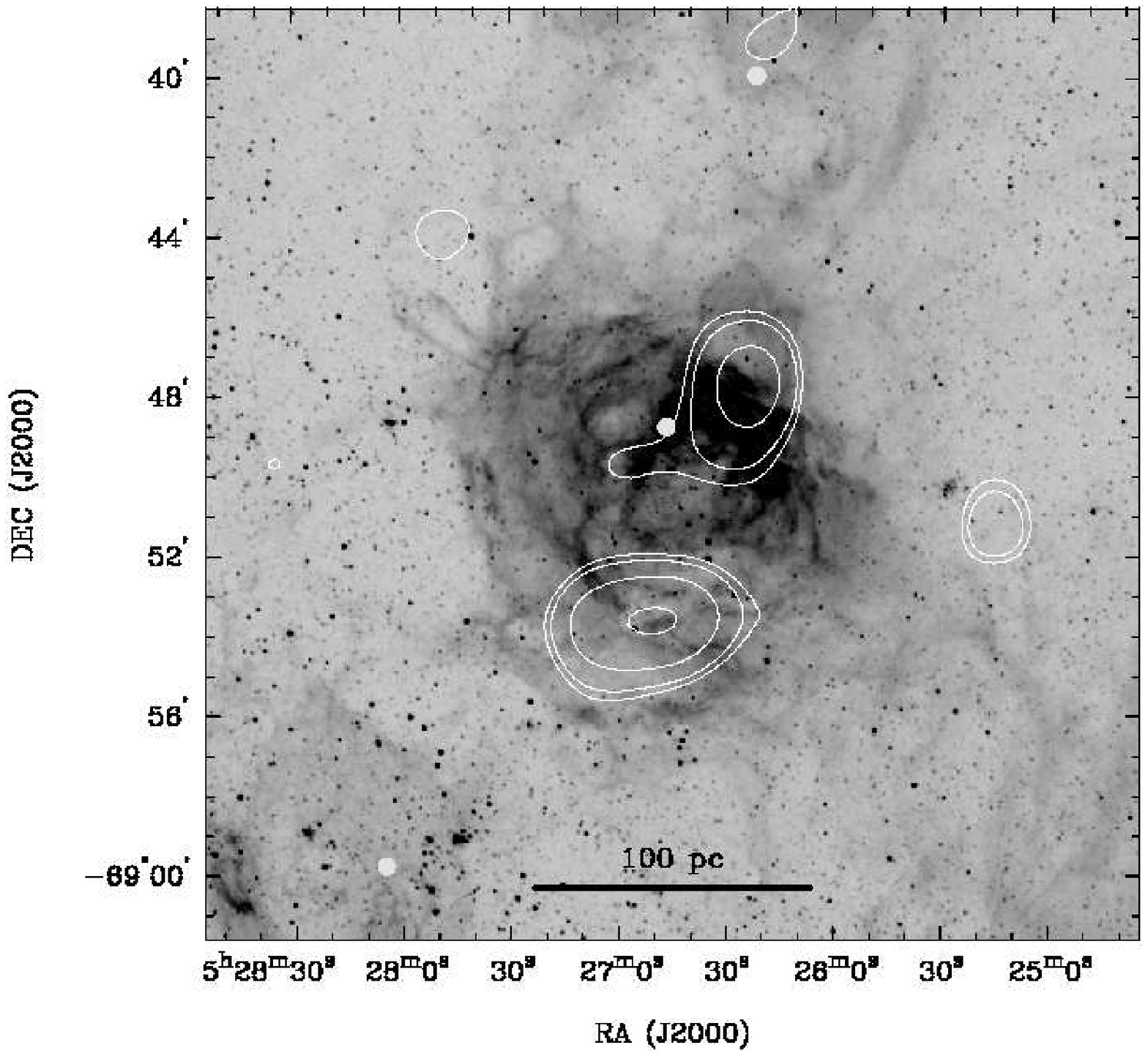,width=2.4in,angle=0}
\psfig{file=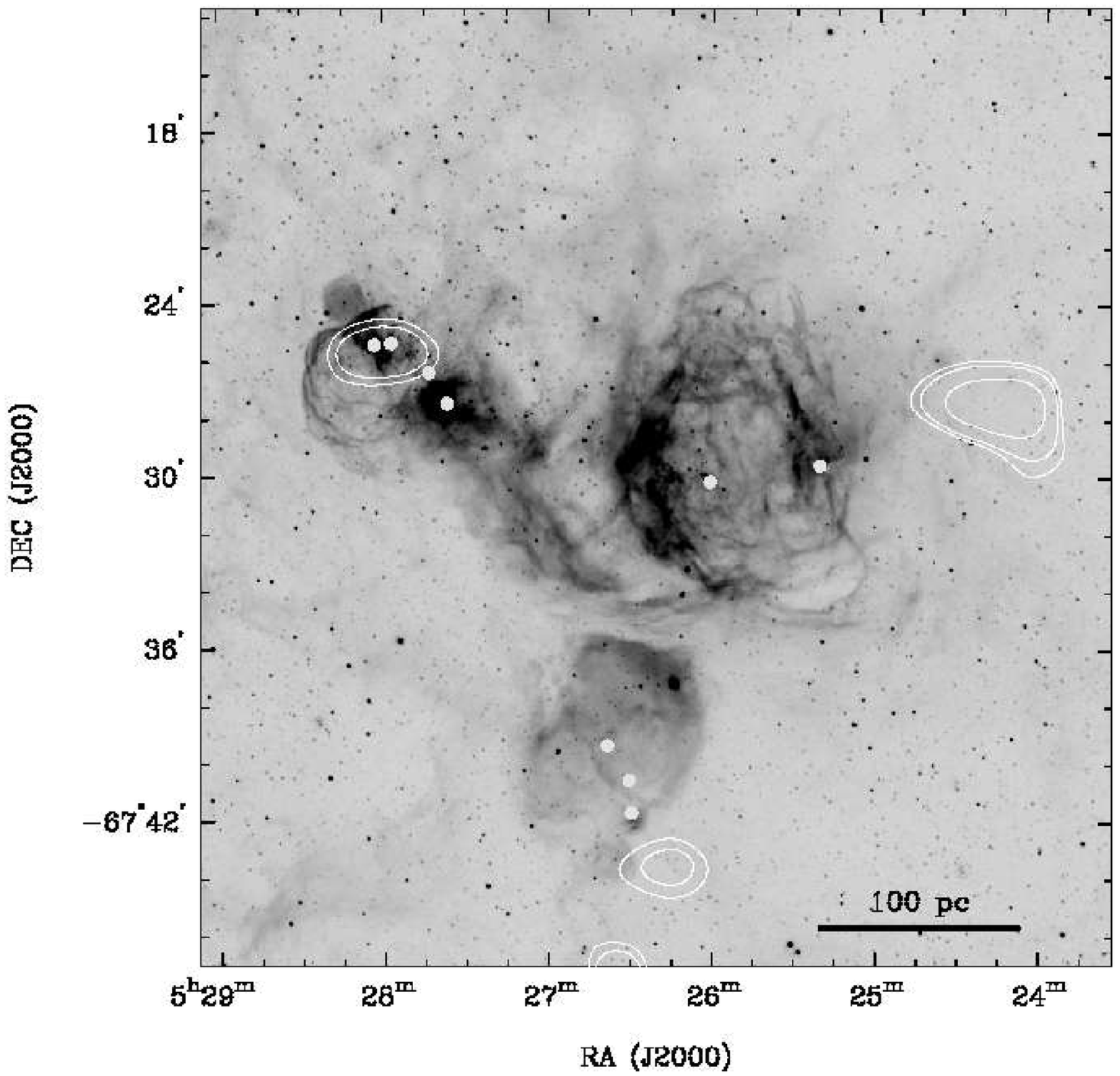,width=2.4in,angle=0}
}
\caption{Massive YSOs in or near OB associations surrounded 
by superbubbles.  The H\,II complexes in the top row are 
N44 (left) and N206 (right), and the bottom row N144 (left)
and N51 (right).  The YSOs are plotted as filled circles
in these MCELS H$\alpha$ images overplotted with CO contours
from the NANTEN survey. 
}
\end{figure}

On a global scale, massive star formation is associated with
gravitational instability.  
The gravitational instability of the LMC has been analyzed
by \citet{Yetal07} considering the gas disk only and considering
both the collisional gas and the collisionless stars in the disk.
They find that only 62\% of the {\it Spitzer} sample of 
massive YSOs fall in gravitationally unstable regions for 
the former, and 85\% for the latter.
The most visible difference between these two cases is along the
star-forming stellar bar of the LMC, which is largely stable for 
the gas-only disk and unstable in the gas+star disk.
It is thus important that the contribution of stars is included in
the calculation of gravitational instability.  

For the 234 probable and 14 possible massive YSOs in the LMC, 
we mark the YSO positions in H$\alpha$ images overplotted with
CO contours to examine their stellar and interstellar environments.
We compare the YSO location with the OB associations compiled by
\citet{LH70}, H\,II regions cataloged by \citet{H56} and \citet{DEM},
and giant molecular clouds identified from NANTEN survey by
\citet{Fetal99}.  The results are compiled in a table for further
statistical analysis.

Among the 248 probable and possible massive YSOs in the LMC,
$\sim$65\% are in OB associations, most of which are surrounded 
by superbubbles or bright
H\,II regions.  Figure 4 shows examples of massive YSOs in or near
OB associations in superbubbles.  It is evident that star formation
continues into the molecular clouds near OB associations, and
most likely the expansion of the superbubble plays a significant
role in the propagation of star formation.  Low-level star formation
exists within the superbubble where no molecular material is detected
by the NANTEN survey; these are probably pc-sized dust globules
compressed by the photoionized surface layer (i.e., photo-implosion
or globule-squeezing), as illustrated by the YSOs in the superbubble
N51D \citep{Cetal05}.

About 85\% of the massive YSOs in the LMC are formed in giant 
molecular clouds.
Interestingly, some massive YSOs are located within giant molecular
clouds but are also on the peripheries of supernova remnants (SNRs).
Figure 5 shows two examples.  While it is possible that the expansion
of the SNR triggered the star formation, it is also possible that
the star formation was triggered by the bubble blown by the
supernova progenitor's fast wind during the main sequence phase and 
the subsequent SNR is largely confined within the bubble.
To distinguish between these two possibilities, velocities of the
SNR shell and the superposed molecular cloud need to be compared.
Bubble expansion velocities are expected to be small, $\le$20 km~s$^{-1}$,
while SNR shocks are expected to be $>$100 km~s$^{-1}$.
An absence of high velocities in the molecular material would suggest
that the expanding wind-blown bubble was responsible for triggering 
the star formation.

\begin{figure}
\centerline{
\psfig{file=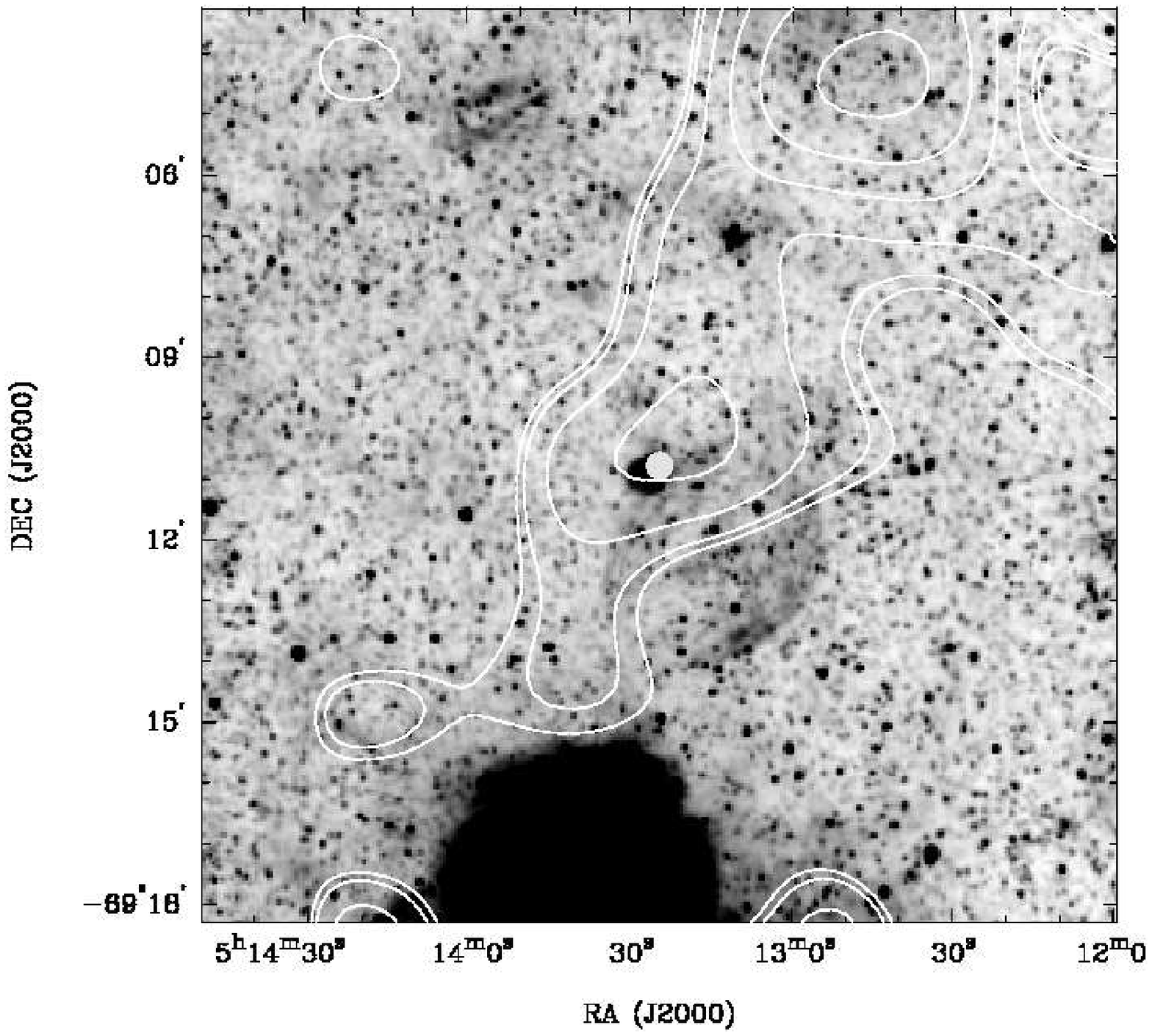,width=2.5in,angle=0}
\psfig{file=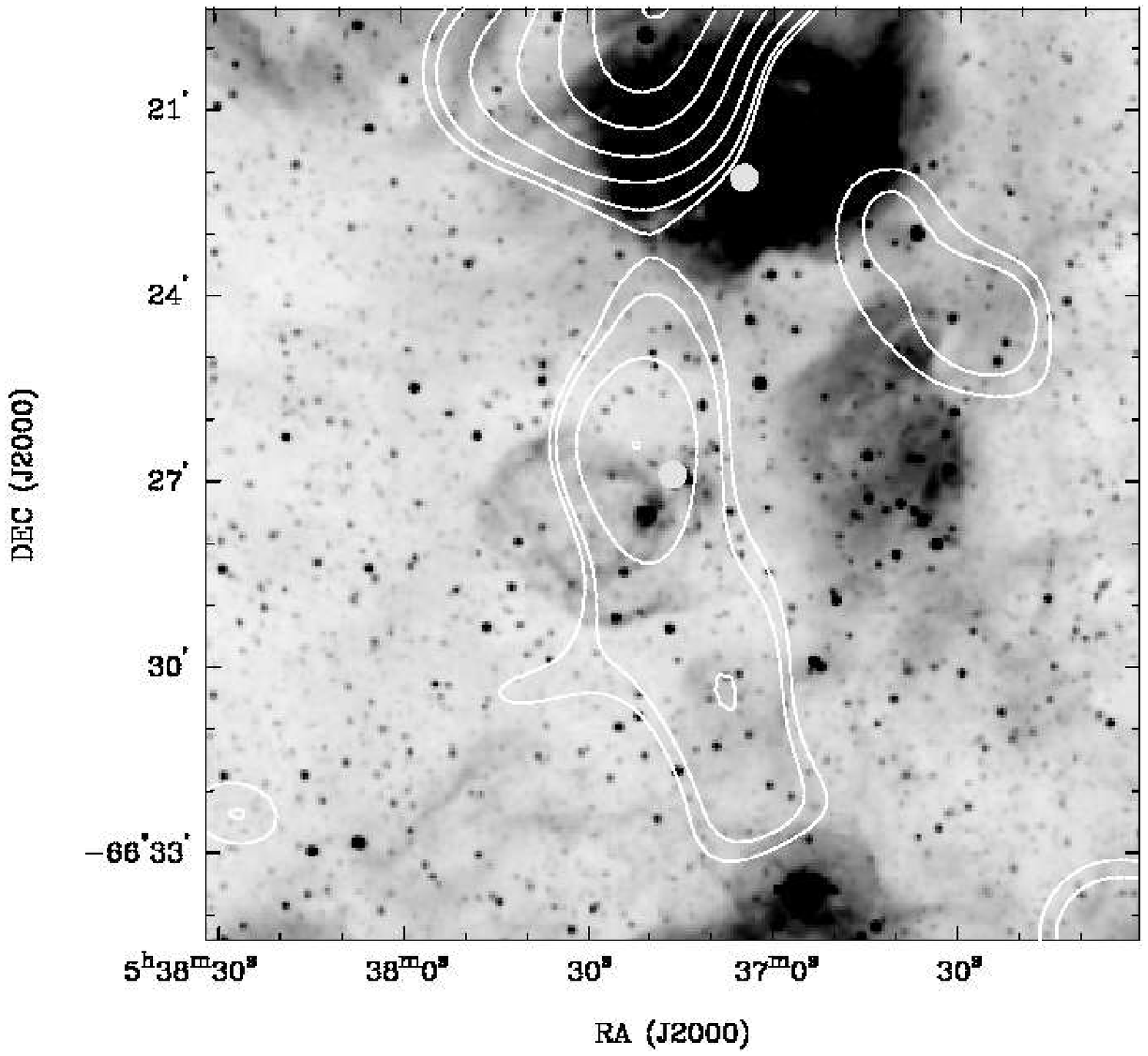,width=2.5in,angle=0} }
\caption{Massive YSOs associated with SNR 0513$-$69.2 (left) 
and SNR DEM\,L256 (right).  The YSOs are plotted as filled
circles in the MCELS H$\alpha$ images overplotted with CO contours 
from the NANTEN survey.  The SNR is the small ($\sim$3$'$ across)
shell near the field center.
}
\end{figure}

\begin{figure}
\centerline{
\psfig{file=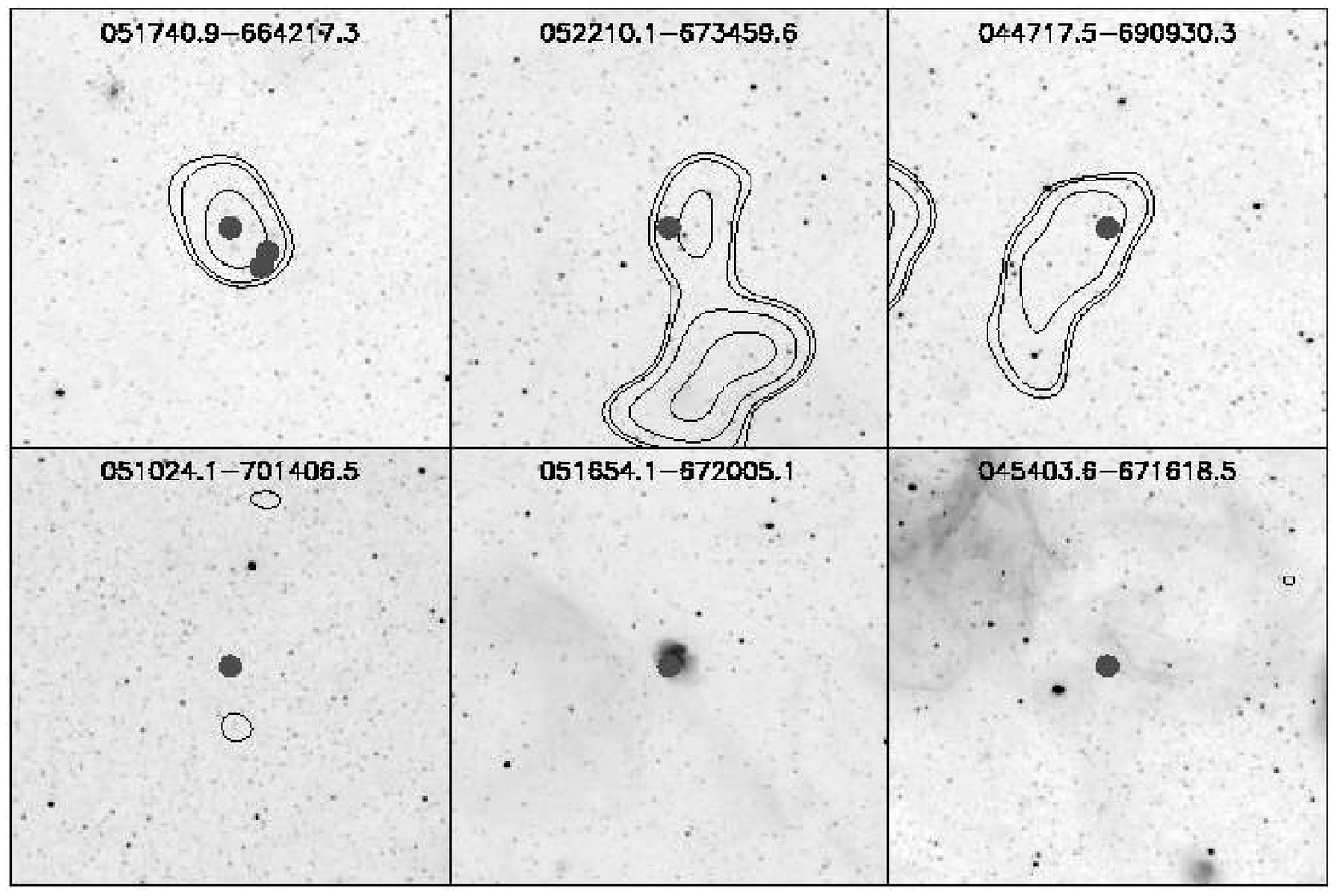,width=5in,angle=0}
}
\caption{Massive YSOs that are not in OB associations or 
crowded environments.  The massive YSOs are plotted as filled
circles in MCELS H$\alpha$ images overplotted with CO
contours from the NANTEN survey.  The field-of-view of each
panel is 15$'$ $\times$ 15$'$, or 225 pc $\times$ 225 pc.
}
\end{figure}


The massive YSOs that are not in OB associations may be
associated with giant molecular clouds, as shown in the
top row of Figure 6.
It is very likely that the stellar energy feedback of these
first-generation massive stars will trigger further star formation,
resulting in an OB association or a cluster.
Only massive YSOs formed in an environment without a large
reservoir of molecular gas may become truly isolated massive stars,
for example, the YSOs shown in the bottom row of Figure 6.
Among the massive YSOs in the LMC, only $\sim$7\% are neither
in OB associations nor associated with giant molcular clouds.
These will evolve into isolated massive stars.
This fraction is consistent with the finding for Galactic 
O-type stars -- only 5-10\% are truly isolated.
It should be noted that the isolated massive YSOs are in general
less luminous, and hence less massive, than those in active
star forming regions.  
Detailed comparisons of YSO properties for different environments
are still underway.

\acknowledgements 
This research is supported by NASA grants JPL 1264494 and 1290956.


\end{document}